\documentclass[sigconf,authorversion,nonacm]{acmart}

\usepackage{booktabs}
\usepackage{graphicx}
\usepackage{colortbl}
\usepackage{xcolor}
\usepackage{enumitem}
\usepackage{multirow}
\usepackage{soul}
\usepackage{xspace}
\usepackage{tikz}
\usepackage{adjustbox}
\usepackage{listings}
\usetikzlibrary{automata, positioning, arrows}
\tikzset{
    ->,  
    >=stealth, 
    node distance=5cm, 
    every state/.style={thick, fill=gray!10}, 
    initial text=$ $, 
    font={\fontsize{20pt}{20}\selectfont},
    }
\definecolor{brinkpink}{rgb}{0.98, 0.38, 0.5}
\definecolor{capri}{rgb}{0.0, 0.75, 1.0}
\definecolor{LightCyan}{rgb}{0.88,1,1}

\newcommand{\alessandro}[1]{\textcolor{red}{{\it [Alessandro: #1]}}}

\newcommand{\changetxt}[1]{\textcolor{black}{#1}\textcolor{black}\xspace}

\newcommand{\adaptive}{\textsc{AdaptiveMon}\xspace}
\newcommand{\static}{\textsc{StaticMon}\xspace}

\newcommand{\adaptRate}{Change Rate\xspace}
\newcommand{\adaptMetrics}{Select Indicators\xspace}
\newcommand{\adaptCombined}{Combined Countermeasures\xspace}

\AtBeginDocument{%
  \providecommand\BibTeX{{%
    \normalfont B\kern-0.5em{\scshape i\kern-0.25em b}\kern-0.8em\TeX}}}

\setcopyright{acmcopyright}
\copyrightyear{2022}
\acmYear{2022}
\setcopyright{acmcopyright}\acmConference[SEAMS '22]{17th International Symposium on Software Engineering for Adaptive and Self-Managing Systems}{May 18--23, 2022}{PITTSBURGH, PA, USA}
\acmBooktitle{17th International Symposium on Software Engineering for Adaptive and Self-Managing Systems (SEAMS '22), May 18--23, 2022, PITTSBURGH, PA, USA}
\acmPrice{15.00}
\acmDOI{10.1145/3524844.3528055}
\acmISBN{978-1-4503-9305-8/22/05}

\begin{document}

\title{Towards Self-Adaptive Peer-to-Peer Monitoring for Fog Environments}

\author{Vera Colombo}
\email{v.colombo51@campus.unimib.it}
\affiliation{%
  \institution{University of Milano-Bicocca}
  \streetaddress{Piazza dell'Ateneo Nuovo, 1}
  \city{Milan}
  \state{Italy}
  \country{IT}
  \postcode{20126}
}

\author{Alessandro Tundo}
\email{alessandro.tundo@unimib.it}
\orcid{0000-0001-8840-8948}
\affiliation{%
  \institution{University of Milano-Bicocca}
  \streetaddress{Piazza dell'Ateneo Nuovo, 1}
  \city{Milan}
  \state{Italy}
  \country{IT}
  \postcode{20126}
}

\author{Michele Ciavotta}
\email{michele.ciavotta@unimib.it}
\orcid{0000-0002-2480-966X}
\affiliation{%
  \institution{University of Milano-Bicocca}
  \streetaddress{Piazza dell'Ateneo Nuovo, 1}
  \city{Milan}
  \state{Italy}
  \country{IT}
  \postcode{20126}
}

\author{Leonardo Mariani}
\email{leonardo.mariani@unimib.it}
\orcid{0000-0001-9527-7042}
\affiliation{%
  \institution{University of Milano-Bicocca}
  \streetaddress{Piazza dell'Ateneo Nuovo, 1}
  \city{Milan}
  \state{Italy}
  \country{IT}
  \postcode{20126}
}

\renewcommand{\shortauthors}{Colombo and Tundo, et al.}

\begin{abstract}
Monitoring is a critical component in fog environments: it promptly provides insights about the behavior of systems, reveals Service Level Agreements (SLAs) violations, enables the autonomous orchestration of services and platforms, calls for the intervention of operators, and triggers self-healing actions. 

In such environments, monitoring solutions have to cope with the heterogeneity of the devices and platforms present in the Fog, the limited resources available at the edge of the network, and the high dynamism of the whole Cloud-to-Thing continuum.

This paper addresses the challenge of accurately and efficiently monitoring the Fog with a self-adaptive peer-to-peer (P2P) monitoring solution that can opportunistically adjust its behavior according to the collected data exploiting a lightweight rule-based expert system. 

Empirical results show that adaptation can improve monitoring accuracy, while reducing network and power consumption at the cost of higher memory consumption.  
\end{abstract}



\maketitle

 
\section{Introduction}\label{sec:introduction}

Fog computing has been proposed to fill the gap between the Cloud and the Internet-of-Things (IoT), providing a continuum of computing, storage, and networking facilities designed to address emerging scenarios related to several domains, such as smart cities, transportation, industry 4.0, and e-health~\cite{bonomi2012fog,hu2017survey,yousefpour2019all}. \changetxt{Its infrastructure is made of a large number of widely distributed heterogeneous nodes, which can include network devices (e.g., routers, gateways, access points, and base stations), micro-clouds and computing servers~\cite{hu2017survey,yousefpour2019all}.}

Monitoring is an intrinsically pivotal component of the Fog as it is the entity responsible for providing reliable information about the behavior of infrastructure and services~\cite{abderrahim2017holistic, aceto2013cloud, taherizadeh2018monitoring}. Monitoring data are used in several tasks, including the autonomous orchestration of services and platforms~\cite{taherizadeh2018monitoring}, the detection of Service Level Agreements (SLAs) violations~\cite{alodib2016qos}, the activation of operator-mediated procedures~\cite{benson2010first}, and self-healing actions~\cite{yousefpour2019all}.

Although a large number of monitoring solutions are already available for the Cloud~\cite{calero2014monpaas, tundo2019varys, trihinas2014jcatascopia, povedano2013dargos, olups2016zabbix,barth2008nagios, turnbull2018monitoring, shatnawi2018cloudhealth,elasticstack_2022}, they are well-known to badly adapt to heterogeneous and massively distributed infrastructures characterized by frequent changes to the topology of the nodes, such as the Fog~\cite{yousefpour2019all}.
For instance, Abderrahim et al.~\cite{abderrahim2017holistic} discuss how a fog monitoring system, unlike cloud-specific solutions, must make particularly good use of the available resources, must be resilient to changes in the topology (e.g., nodes joining and leaving the network) and network conditions (e.g., communication links may not always be fully operational).


\emph{Peer-to-Peer (P2P) architectures} have been recently investigated as viable approaches to effectively address monitoring in the Fog~\cite{yousefpour2019all,abderrahim2017holistic,graffi2017skyeye, forti2021lightweight}. 
Indeed, as they are ``self-organizing systems of equal, autonomous entities (peers) which aim for the shared usage of distributed resources in a networked environment avoiding central services"~\cite{oram2001peer}, they represent a legitimate option to address the dynamism of the Fog without imposing strong constraints on the stability of the target environment.
For instance, FogMon~\cite{forti2021lightweight} is a P2P monitoring solution targeting fog infrastructures that can monitor hardware resources, end-to-end network QoS, and connected IoT devices. SkyEye~\cite{graffi2017skyeye} is a P2P tree-based monitoring approach that provides lightweight continuous monitoring of all peers in its network. 

Unfortunately, although these monitoring systems show some degree of adaptivity thanks to the features provided by P2P architectures (e.g., they can tolerate node disconnections and broken communication links), they lack adaptation mechanisms that take into account the monitored indicators~\cite{forti2021lightweight, abderrahim2017holistic, graffi2017skyeye}. In fact, the collected indicators reveal important information about the monitored system and its environment, and can be exploited to increase the awareness and adaptability of the monitoring system itself. For example, a monitoring node running in a device exhausting its battery may stop monitoring the non-essential indicators. Similarly, the trend of a monitored indicator can be used to optimize the sampling rate to avoid wasting resources (e.g., increasing/decreasing the sampling rate based on the degree of stability of the indicator).


In this paper, we address this limitation by proposing a \textit{self-adaptive P2P monitoring} solution that exploits a hierarchical P2P architecture and embeds adaptive behaviors defined according to the MAPE-K feedback loop~\cite{kephart2003vision}. The proposed monitoring system can abstract the monitored indicators and activate countermeasures based on the status of these indicators. Countermeasures are defined by exploiting a lightweight rule-based system embedded in the peers. In the paper, we show how this framework can be used to activate two example countermeasures that can dynamically change the set of monitored indicators and their sampling rate. 

We implemented our solution as an extension of FogMon~\cite{forti2021lightweight}, a state-of-art P2P monitoring solution for the Fog, and released it publicly\footnote{https://github.com/veracoo/FogMon/tree/adaptive-fogmon}. The empirical evaluation of the accuracy and effectiveness of the adaptive version of the monitoring solution in contrast with the non-adaptive one shows that adaptive behaviors can be used to increase the accuracy of the collected data and save network and power consumption at the cost of higher memory consumption. This is an interesting trade-off since devices are often reasonably equipped in terms of memory, while network bandwidth and power are critical resources at the edge~\cite{abderrahim2017holistic}. 

In a nutshell, the main contributions of this paper are:

\begin{itemize}[leftmargin=*]
    \item A self-adaptive monitoring solution that modifies its behavior by running countermeasures defined over an abstraction of the monitored indicators.
    \item The instantiation of the MAPE-K loop on the case of P2P fog monitoring.
    \item The definition of two countermeasures to self-adapt the monitoring system: (i) \emph{\adaptRate}, which can be used to dynamically alter the data collection rate; and (ii) \emph{\adaptMetrics}, which can be used to dynamically control the activation and deactivation of the probes.
    \item A publicly available prototype obtained by extending the FogMon P2P monitoring system.
    \item An empirical evaluation of the monitoring accuracy and efficiency of the adaptive solution in comparison to the non-adaptive counterpart.
\end{itemize}

The rest of this paper is organized as follows. Section~\ref{sec:p2p-monitoring} discusses the use of P2P architectures for monitoring in the Fog. Section~\ref{sec:approach} provides a rigorous definition of the self-adaptive P2P monitoring solution proposed in this paper. Section~\ref{sec:prototype} briefly describes our prototype implementation. Section~\ref{sec:evaluation} presents the empirical results about monitoring accuracy and efficiency. Section~\ref{sec:related-work} discusses related work. Finally, Section~\ref{sec:conclusions} provides final remarks.
 
\section{P2P Monitoring}\label{sec:p2p-monitoring}
A P2P architecture consists of a network of autonomous self-\linebreak organizing entities (i.e., peers) that employ distributed resources to accomplish a common task in a decentralized fashion, thus, without relying on central services~\cite{oram2001peer,milojicic2002peer,steinmetz2005peer}.

The P2P architecture provides the applications with the capability to deal with some of the highly dynamic traits of fog computing, increasing the tolerance to both network failures and nodes joining and leaving the system~\cite{abderrahim2017holistic, forti2021lightweight, Ciavotta2021DFaas}. 
Furthermore, it provides autonomy, scale, and robustness, which are critical capabilities to operate in such an environment~\cite{tracey2019see}. 
Finally, P2P architectures make monitoring data available across the network without relying on a single centralized component, but rather on a set of peers 
constituting a self-organized overlay network.
This is particularly beneficial when the connectivity to the Cloud is limited, such as during disasters or severe network outages~\cite{yousefpour2019all}.

\begin{figure}[t]
    \centering
    \includegraphics[width=0.9\linewidth]{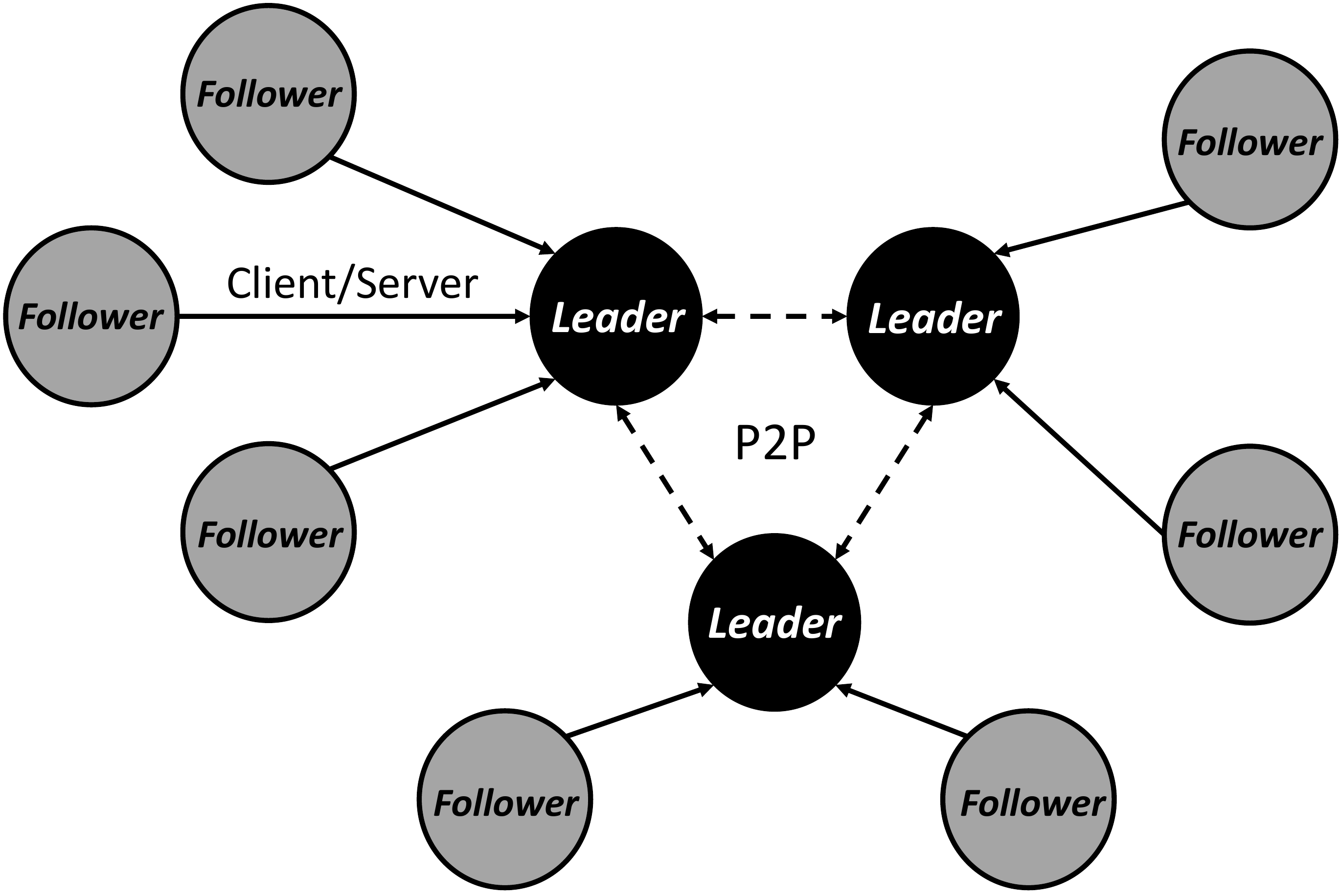}
    \caption{Hierarchical P2P monitoring architecture proposed by Forti et al.~\cite{forti2021lightweight}.}
    \label{fig:fogmon-architecture}
\end{figure}

In this work, we use the two-tier hierarchical P2P monitoring architecture~\cite{yang2003designing, kleis2005hierarchical,forti2021lightweight} proposed by Forti et al.~\cite{forti2021lightweight} that is shown in Figure~\ref{fig:fogmon-architecture}, with \textit{Followers} at the lower tier and \textit{Leaders} at the higher tier.

The benefit of employing such an architecture in the Fog is twofold. First, it implies different roles for peers running in different tiers, depending on the available resources. Followers are designed to collect data by running on the edge, within nodes and devices with limited resources. Leaders are designed to consume more resources to store the data received from the Followers while creating and operating the P2P network. Followers are connected to a single Leader and work in a classic client-server fashion~\cite{yang2003designing}. 
These distinct roles can be used to opportunistically exploit the available resources, including the possibility to adapt to changing conditions (e.g., bandwidth or resource degradation) through dynamic peer promotion/demotion.

Second, it helps reducing the network overhead by limiting the amount of data transferred between the peers. Actually, Followers can forward data to their Leader only, leaving the thinner upper-tier with the responsibility of building a global state of the monitored resources by exchanging monitoring data among Leaders.

We refer to FogMon as reference implementation for this architecture~\cite{forti2021lightweight}. 
In FogMon, the Followers monitor their own deployment node by probing hardware resources (i.e., CPU, memory, and hard disk), collecting end-to-end network QoS data (i.e., latency and bandwidth), and detecting IoT devices. 
Data is collected and sent to Leader nodes at a \textit{fixed rate}. To limit network overhead, Followers send differential updates, that is, they only send data whose average or variance differ more than a \emph{sensitivity} threshold (i.e., 10\% by default) from the last value sent~\cite{forti2021lightweight}.
Leaders periodically aggregate \changetxt{monitoring} data received from Followers, and share the aggregated data with the other Leaders through a gossip protocol~\cite{jelasity2011gossip}.

\section{Self-Adaptive P2P Monitoring} \label{sec:approach}

\begin{figure}[t]
    \centering
    \includegraphics[width=0.6\linewidth]{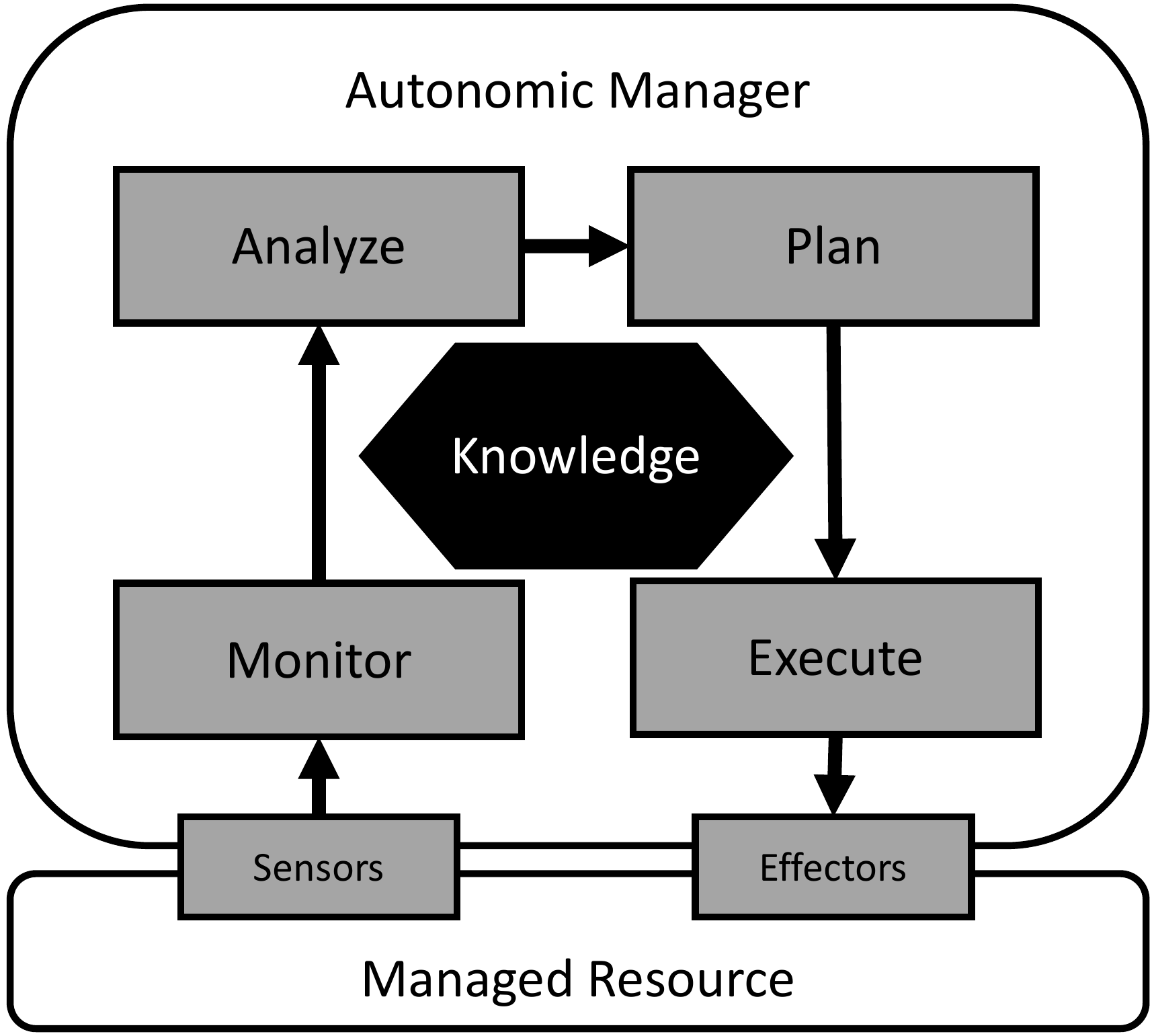}
    \caption{Monitor, Analyze, Plan, Execute, and Knowledge (MAPE-K) loop as proposed by Kephart and Chess~\cite{kephart2003vision}.}
    \label{fig:mape-k-loop}
\end{figure}

This section describes how P2P monitoring can be enhanced with self-adaptive capabilities to both make a better use of the available resources and enable the capability to promptly react to run-time events. We refer to the self-adaptive version of the P2P monitoring solution presented in this paper as \adaptive, in contrast with the non-adaptive version that we refer to as \static. 

\adaptive enriches the capabilities of the monitoring system by embedding the Monitor, Analyze, Plan, Execute, and Knowledge (MAPE-K) control framework~\cite{kephart2003vision}, shown in Figure~\ref{fig:mape-k-loop}, within each peer.  
The \emph{monitor} component of the MAPE-K loop collects data about a managed resource through \emph{sensors}. 
In our case, this is achieved by the monitoring probes running within the peers. The \emph{analyze} and \emph{plan} steps analyze the collected data and plan for the countermeasures to be activated. Finally, the \emph{execute} step exploits \emph{effectors} to run the selected countermeasures. In our case, the countermeasures reconfigure the monitoring systems according to the observations collected from the managed resource. The \emph{knowledge} about the managed resource is shared among all the components. 
\changetxt{Note that the MAPE-K loop runs withing each peer, independently of the overall architecture, which gives peers the capability to run self-adaptive behaviors regardless of their role within the architecture.}

In the following, we describe how the components of the MAPE-K loop embedded in the peers are defined, and present two countermeasures that we experienced in our prototype implementation, namely (i) \emph{\adaptRate}, which adjusts the rate Followers sample and forward data to their Leader, and (ii) \emph{\adaptMetrics}, which dynamically activates and deactivates the set of monitored indicators. 

\subsection{Knowledge}

\begin{table*}[!t]
\centering
\begin{adjustbox}{width=0.7\linewidth}
\addtolength{\tabcolsep}{-2pt}
\begin{tabular}{@{}l|l|ll@{}}
\toprule
\textbf{Indicator} &
  \textbf{State} &
  \textbf{Definition} \\ 
  \textbf{Type} & &\\ 
  \midrule
\multirow{2}{*}{\textit{Categorical}} &
  stable &
  $s_t = \textit{stable} \iff v_t = v_{t-x} \quad \forall x \in [1, k]$ \\ \cmidrule{2-3}
 &
  unstable &
  $s_t = \textit{unstable} \iff \exists x \in [1, k] \mid v_t \neq v_{t-x}$ \\ \midrule
\multirow{15}{*}{\textit{Numerical}} &
  stable &
  \begin{tabular}[c]{@{}l@{}}$s_t = \textit{stable} \iff |Stab_t| \geq p \cdot k \land |v_{t} - v_{t-1}| \leq \Delta_{max}$,\\ $Stab_t = \{|v_{x} - v_{x-1}| \leq \Delta_{max}\}_{x \in [1, k]}$\end{tabular} \\ \cmidrule{2-3}
 &
  unstable &
  \begin{tabular}[c]{@{}l@{}}$s_t = \textit{unstable} \iff |Stab_t| < p \cdot k \lor |v_{t} - v_{t-1}| > \Delta_{max}$,\\ $Stab_t = \{|v_{x} - v_{x-1}| < \Delta_{max}\}_{x \in [1, k]}$\end{tabular} \\ \cmidrule{2-3}
 &
  too high &
  \begin{tabular}[c]{@{}l@{}}$s_t = \textit{too high} \iff |Too\_High_t| \geq p \cdot k \land v_{t} \in I_\textit{too\_high}$,\\ $Too\_High_t = \{v_x \in I_\textit{too\_high}\}_{x \in [0, k]}$\end{tabular} \\ \cmidrule{2-3}
 &
  high &
  \begin{tabular}[c]{@{}l@{}}$s_t = \textit{high} \iff |High_t| \geq p \cdot k \land v_{t} \in I_\textit{high}$,\\ $High_t = \{v_x \in I_\textit{high}\}_{x \in [0, k]}$\end{tabular} \\ \cmidrule{2-3}
 &
  normal &
  \begin{tabular}[c]{@{}l@{}}$s_t = \textit{normal} \iff |Normal_t| \geq p \cdot k \land v_{t} \in I_\textit{normal}$,\\ $Normal_t = \{v_x \in I_\textit{normal}\}_{x \in [0, k]}$\end{tabular} \\ \cmidrule{2-3}
 &
  low &
  \begin{tabular}[c]{@{}l@{}}$s_t = \textit{low} \iff |Low_t| \geq p \cdot k \land v_{t} \in I_\textit{low}$,\\ $Low_t = \{v_x \in I_\textit{low}\}_{x \in [0, k]}$\end{tabular} \\ \cmidrule{2-3}
 &
  too low &
  \begin{tabular}[c]{@{}l@{}}$s_t = \textit{too low} \iff |Too\_Low_t| \geq p \cdot k \land v_{t} \in I_\textit{too\_low}$,\\ $Too\_Low_t = \{v_x \in I_\textit{too\_low}\}_{x \in [0, k]}$\end{tabular} \\ \bottomrule[1pt]
  \toprule
  \multicolumn{2}{l|}{\textbf{Symbols}} & \textbf{Definition}\\
  \midrule
   \multicolumn{2}{l|}{$||$} &
  Cardinality of a set. \\  \midrule
   \multicolumn{2}{l|}{$k$} &
  Number of samples considered in the recent history of an indicator. \\  \midrule
 \multicolumn{2}{l|}{$p \in [0, 1]$} &
  Tolerance parameter that indicates the percentage of $k$ samples.\\
  \multicolumn{2}{l|}{}&
  It must satisfy the constraint that characterizes the state definition. \\ \midrule
 \multicolumn{2}{l|}{$\Delta_{max}$} & 
 Maximum delta allowed to consider an indicator as stable.
   \\ \midrule
\multicolumn{2}{l|}{$I_\textit{too\_high} = [\textit{too\_high}, +\infty)$} & Interval of indicator values considered \textit{too high}.
   \\ \midrule
    \multicolumn{2}{l|}{$I_\textit{high} = [\textit{high}, \textit{too\_high})$} & Interval of indicator values considered \textit{high}.
   \\ \midrule
 \multicolumn{2}{l|}{$I_\textit{normal} = (\textit{low}, \textit{high} )$} & Interval of indicator values considered \textit{normal}.
   \\ \midrule
 \multicolumn{2}{l|}{$I_\textit{low} = (\textit{too\_low}, \textit{low} ]$} & Interval of indicator values considered \textit{low}.
   \\ \midrule
 \multicolumn{2}{l|}{$I_\textit{too\_low} = (-\infty, \textit{too\_low}] $} & Interval of indicator values considered \textit{too low}. \\ \bottomrule
\end{tabular}
\end{adjustbox}
\caption{States definitions for categorical and numerical indicators.}
\label{tab:metrics-states}
\end{table*}

The knowledge exploited in \adaptive{} consists of the monitored \emph{indicators}, which represent the raw knowledge about the monitored resource, and the associated \emph{logical states}, which capture the semantics of the values of an indicator.

\begin{definition}[Indicator Values and time series]\label{def:metric-value}
Given a monitored indicator $I$ and a domain $D$ of values for $I$, $v^{I}_{t} \in D$ denotes the value of the indicator $I$ at time $t$. A sequence of values for a same indicator generates a time series, that is,  $v^I_i, v^I_{i+1}, \ldots, v^I_k$ is a time series for indicator $I$.
\end{definition}

\begin{definition}[Logical States]\label{def:metric-state}
Given a monitored indicator $I$ and a finite set of abstract states $S$, $S^{I}_{t} \subseteq S$ represents a potentially empty set of logical states associated with the indicator $I$ at time $t$. The sequence of states sets $S^I_i, S^I_{i+1}, \ldots, S^I_k$ associated with an indicator also forms a time series.
\end{definition}

For sake of notation, we omit $I$ when the indicator is obvious from the context.

While time series of values simply reflect the sequence of probed samples, the corresponding time series of logical states captures the state of an indicator at a specific time, revealing information about the monitored resource. For instance, an indicator might be unstable, too high, or within a normal range of values. These states can be derived from the time series of raw values and used to fire countermeasures, as explained below. 

We defined a set of logical states useful for our countermeasures, but this set can be clearly extended depending on the countermeasures to be implemented.
We summarize the rigorous definitions of the logical states in Table~\ref{tab:metrics-states}, while we discuss them informally below. 

In case the domain of a metric is \emph{categorical}, we defined two logical states: \emph{stable} and \emph{unstable}. A categorical indicator is stable at a time $t$ if its value has been constant in the recent history of the execution, otherwise it is unstable.



In case the domain of an indicator is \emph{numerical}, we defined seven states: \emph{stable}, \emph{unstable}, 
\emph{normal}, \emph{high}, \emph{low}, \emph{too high}, \emph{too low}. 
A numerical indicator is stable at a time $t$ if its value is close to its value at time $t-1$, and the indicator had small variations in the recent history of the execution. Similarly, a numerical indicator is unstable at a time $t$ if its value differs significantly from the value at time $t-1$, and the indicator had significant variations in the recent history of the execution.



The remaining five states detect values that steadily stay in a given region of the domain. In particular, an indicator is too high or high if its value is above given thresholds and the indicator has been mostly above those thresholds in its recent execution history. 
Similarly, an indicator is too low or low if its value is below given thresholds and the indicator has been mostly below those thresholds in its recent execution history. 
Finally, an indicator is normal if its value is in the normal range, and the indicator has been mostly normal in its recent history of the execution. The thresholds for the various levels are defined on a per-indicator basis since they depend on both the indicator and the application domain. For instance, threshold values for memory and CPU consumption are clearly different, and threshold values of memory consumption for a memory-intensive application and a lightweight application are also different.

The set of the collected indicators, along with their raw values and state values represent the knowledge available to \adaptive.

\subsection{Monitor}
Monitoring is rather natural and inexpensive in \adaptive since Followers collect data from a monitored resource by construction, and thus the same data sent to Leaders is also available to the MAPE-K loop. If needed, extra indicators can be collected for the only purpose of controlling the adaptive behavior of the peers, even if not needed by the applications accessing the data produced by the monitoring system.

The monitoring behavior is controlled by a \emph{sampling rate} parameter that determines how frequently values $v^I_t$ are collected and forwarded to Leader peers.



\subsection{Analyze}
The analysis mainly consists of a data processing routine that converts the raw values collected for every indicator into its logical state representation. In particular, the analysis process accesses the collected values and applies the definitions reported in Table~\ref{tab:metrics-states} to generate a time series of logical states for every monitored indicator. 

\begin{figure}[!ht]
    \centering
    \resizebox{\linewidth}{!}{%
    \begin{tikzpicture}
        \coordinate(c6) at (-2, 4.8);
        \coordinate(c7) at (-1.5, 5.1);
        \coordinate(c8) at (-1, 5.2);
        \coordinate(c9) at (0, 5.5);
        \coordinate(c10) at (0.5, 5.6);
        \coordinate(c11) at (1, 5.6);
        \coordinate(c12) at (1.5, 5.7);
        \coordinate(c13) at (2, 5.6);
        \coordinate(c14) at (2.5, 5.6);
        \coordinate(c15) at (3, 5.4);
        \coordinate(c16) at (3.5, 5.2);
        
        \coordinate(c17) at (4, 4.5);
        \coordinate(c18) at (4.5, 0.7);
        \coordinate(c19) at (5, 0.3);
        \coordinate(c20) at (5.5, 0.5);
        \coordinate(c21) at (6, 0.4);
        \coordinate(c22) at (6.5, 0.8);
        
        \coordinate(c23) at (7, 3.5);
        \coordinate(c24) at (7.1, 2);
        \coordinate(c25) at (7.2, 3);
        \coordinate(c26) at (7.5, 3.3);
        \coordinate(c27) at (8, 3.5);
        \coordinate(c28) at (8.5, 3.6);
        \coordinate(c29) at (9, 3.7);
        \coordinate(c30) at (9.5, 3.5);
        \coordinate(c31) at (10, 3.3);
        \coordinate(c32) at (10.5, 3.5);
        \coordinate(c33) at (11, 3.4);
        \coordinate(c34) at (11.5, 3.3);
        
        \coordinate(c35) at (12.0, 3.1);
        \coordinate(c36) at (12.5, 4);
        \coordinate(c37) at (13, 3);
        \coordinate(c38) at (13.5, 6.5);
        \coordinate(c39) at (14, 2);
        \coordinate(c40) at (14.5, 4);
        \coordinate(c41) at (15, 2);

    \draw [-,ultra thick] (c6) to (c7) to (c8) to (c9) to (c10) to (c11) to (c12) to (c13) to (c14) to (c15) to (c16) to (c17) to (c18) to (c19) to (c20) to (c21) to (c22) to (c23) to (c24) to (c25) to (c26) to (c27) to (c28) to (c29) to (c30) to (c31) to (c32) to (c33) to (c34) to (c35) to (c36) to (c37) to (c38) to (c39) to (c40) to (c41);
    
        \draw[->,thick] (-3,0)--(16,0) node[right]{$t$};
        \draw[->,thick] (-2,-1)--(-2,7.5) node[above]{$v^I$};
        \draw[-, thin,orange] (-3, 1)--(16, 1) node[right]{\textit{too low}};
        \draw[-, thin,orange] (-3, 6)--(16, 6) node[right]{\textit{too high}};
        \draw[-, thin,violet] (-3, 1.7)--(16, 1.7) node[right]{\textit{low}};
        \draw[-, thin,violet] (-3, 5.3)--(16, 5.3) node[right]{\textit{high}};
        
        \draw[-, thin,capri,dashed] (-1, -1.5)--(-1, 7) node[]{};
        \draw[-, thin,brinkpink,dashed] (0.5, -2.5)--(0.5, 7) node[]{};
        \draw[-, thin,brinkpink,dashed] (3, -2.5)--(3, 7) node[]{};
        \draw[-, thin,capri,dashed] (3.5, -1.5)--(3.5, 7) node[]{};
        
        \draw[-, thin,capri,dashed] (4.5, -1.5)--(4.5, 7) node[]{};
        
        \draw[-, thin,brinkpink,dashed] (4.5, -2.5)--(4.5, 7) node[]{};
        \draw[-, thin,brinkpink,dashed] (6.5, -2.5)--(6.5, 7) node[]{};
        
        \draw[-, thin,capri,dashed] (7.5, -1.5)--(7.5, 7) node[]{};
        
        \draw[-, thin,brinkpink,dashed] (7.5, -2.5)--(7.5, 7) node[]{};
        \draw[-, thin,brinkpink,dashed] (12.5, -2.5)--(12.5, 7) node[]{};  
        
        \draw[-, thin,capri,dashed] (8, -1.5)--(8, 7) node[]{};
        \draw[-, thin,capri,dashed] (12, -1.5)--(12, 7) node[]{};
        
        \draw[-, thin,capri,dashed] (12.5, -1.5)--(12.5, 7) node[]{};
        \draw[-, thin,capri,dashed] (15, -1.5)--(15, 7) node[]{};
        
         \draw[<->, thin] (-1, -1)--(3.5, -1) node[pos=0.5, above]{\textit{stable}};
         \draw[<->, thin] (0.5, -2)--(3, -2) node[pos=0.5, above]{\textit{high}};
         \draw[<->, thin] (4.5, -1)--(7.5, -1) node[pos=0.5, above]{\textit{unstable}};
         \draw[<->, thin] (4.5, -2)--(6.5, -2) node[pos=0.5, above]{\textit{too low}};
         \draw[<->, thin] (7.5, -2)--(12.5, -2) node[pos=0.5, above]{\textit{normal}};
         \draw[<->, thin] (8, -1)--(12, -1) node[pos=0.5, above]{\textit{stable}};
         \draw[<->, thin] (12.5, -1)--(15, -1) node[pos=0.5, above]{\textit{unstable}};

    \end{tikzpicture}
    }
\caption{
\changetxt{An example of the computed states with respect to the time series values at different time instants.}}
\label{fig:metric-states}
\end{figure}
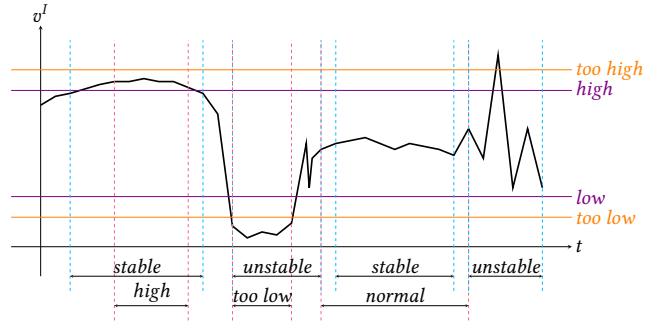

Figure~\ref{fig:metric-states} visually exemplifies the logical states that can be associated with a time series, according to the definitions in Table~\ref{tab:metrics-states}. The logical states are represented as annotations on the $X$-axis. Note that in the example, depending on the shape of the curve, a same point may have 0, 1 or up to 2 logical states associated.

\subsection{Plan}
\adaptive embeds the lightweight CLIPS expert system~\cite{riley1991clips}, which is responsible for determining the countermeasures that have to activate according to the accumulated knowledge based on a set of adaptation rules. 

An adaptation rule consists of two parts: an \textit{antecedent}, that is a set of conditions on the logical states of the indicators that must be satisfied to fire the rule, and a \textit{consequent}, that is a countermeasure to be executed. 

We extended CLIPS by implementing functions that can be used to evaluate the logical states of the indicators, such as functions that can check specific conditions on the last few samples of an indicator. These functions can be used as part of the adaptation rules specified using the CLIPS Domain-Specific Language (DSL). 
For instance,  Listing~\ref{lst:rule} shows an adaptation rule defined to fire the \emph{\adaptRate} countermeasure when the CPU consumption has a stable trend. The symbol \texttt{=>} separates the antecedent and the consequent of the rule. 

It is worth noting that the example adaptation rule uses some of the functions we defined in \adaptive. 
It checks if the CPU consumption is in a stable state with the  \texttt{is\_indicator\_in\_state} function and it computes the new rate for such indicator by executing the \texttt{compute\_indicator\_rate} function. 
The new rate value is in turn used, along with other variables retrieved from the knowledge base, by the \texttt{change\_rate} function that implements the \emph{\adaptRate} countermeasure.

During the plan phase, \adaptive \changetxt{uses the CLIPS expert system to take adaptation decisions, that is, CLIPS} evaluates the antecedents of every rule and inserts the countermeasures activated by the consequent of the fired rules in a priority-based queue. 
\changetxt{Also, }CLIPS handles the activation of the rules by preventing their simultaneous execution.

To demonstrate the self-adaptive capabilities of \adaptive we defined two specific countermeasures: \emph{\adaptMetrics} and \emph{\adaptRate}. 

The \emph{\adaptMetrics} countermeasure can change the set of monitored indicators, either activating or deactivating some of them. The \emph{\adaptRate} countermeasure changes the rate used to sample and send data to Leaders based on the current trends of the indicators. 
The goal of the countermeasure is to gradually increase (decrease) the rate while the monitored indicator is less (more) stable. In particular, the countermeasure updates the sampling rate of the indicator $I$ within predetermined boundaries proportionally to the number of consecutive samples with the same logical state out of the last $k$ samples collected.

We exploited these countermeasures in the context of several adaptation rules. For instance, we defined two rules that can enable/disable monitoring for every indicator different from power consumption if the power is above/below a given threshold, to limit the chance a battery-powered device is abruptly shut down. We also defined two of rules to adapt the sampling and forwarding rate of CPU indicator to its trend.

\begin{lstlisting}[caption=\changetxt{An example} rule that uses the \emph{\adaptRate} countermeasure \changetxt{written with the CLIPS DSL}. \changetxt{The symbol \texttt{=>} separates the antecedent and the consequent of the rule. The \texttt{salience} value represents the rule priority. The \texttt{bind} operator assigns the result of a function call to a variable.},label=lst:rule,basicstyle=\ttfamily\footnotesize,breaklines=true]

(defrule adapt_cpu_rate_when_stable (declare (salience 10))
 (is_indicator_in_state (indicator cpu) (state stable))
 (has_parameter (rate ?current_rate))
    =>
 (bind ?num_of_states (count_indicator_states_in "cpu" "stable"))
 (change_rate "cpu" (compute_indicator_rate "stable" ?num_of_states ?current_rate))
)
\end{lstlisting}

\subsection{Execute}
In this phase, \adaptive{} merely executes countermeasures by running their implementation according to their priority of activation. The actual countermeasures we defined act on the configuration of the peers adapting their behavior to the evolution of the indicators. The actuation interface is straightforward since a peer can directly access the internal variables that govern its behavior.
 
\section{Prototype}\label{sec:prototype}
We implemented \adaptive{} by extending the open-source \changetxt{C++} FogMon P2P monitoring tool~\cite{forti2021lightweight} along multiple dimensions.

In particular, (i) we extended the structure of the peer's local storage \changetxt{(based on the SQLite\footnote{https://www.sqlite.org} database)} to store the logical states used to classify of the monitored indicators; (ii) we embedded the CLIPS rule-based expert system~\cite{riley1991clips} to support the implementation of adaptation rules; (iii) we extended the peers to incorporate adaptive behaviors 
; (iv) we added \changetxt{helper} functions that can be used as part of the adaptation rules to interact with the knowledge; and (v) we implemented the \emph{\adaptMetrics} and \emph{\adaptRate} countermeasures 
\changetxt{to dynamically change the set of the collected indicators and the sampling rate parameters}.
We have not extended the set of monitored indicators, since the indicators already collected by FogMon to monitor the environment were already sufficient to control the activation of our countermeasures. 

The resulting prototype is publicly available with an open-source licence at  \url{https://github.com/veracoo/FogMon/tree/adaptive-fogmon}.

\section{Evaluation}\label{sec:evaluation}
To assess \adaptive{}, we investigate the following two research questions about its effectiveness (monitoring accuracy) and efficiency (resource consumption). 

\subsection{Research Questions}
\noindent \textbf{RQ1 (Monitoring Accuracy) - Can  \adaptive improve the monitoring accuracy of \static?} This research question investigates whether the \emph{\adaptRate} policy of \adaptive can provide a better monitoring accuracy than \static, considering multiple representative trends of the monitored indicators. 

\noindent \textbf{RQ2 (Resource Consumption) - Can  \adaptive save node resources compared to \static?} This research question studies whether the adaptive behavior of \adaptive reduces resource utilization compared to \static. 
We assess the impact of the \emph{\adaptRate} and \emph{\adaptMetrics} countermeasures on resource consumption, both individually and jointly. 


We run all the experiments on a Linux virtual machine (Intel i7-9700 CPU @ 3.00GHz x 4, 4 GB RAM, 13 GB SSD, Ubuntu 20.04 LTS 64-bit, Docker v20.10.0). All the peers run inside dedicated Docker containers, deployed on the same host, and communicate over a bridged network. 
We limit the computational and network resources of the container executing the Follower agent to reproduce a scenario involving resource-constrained and battery-powered devices.
Reference devices are single-board computers (SBC) and micro-controller units (MCU)~\cite{johnston2018commodity, dobrilovic_flammini_gaglione_tokody_2021, seedstudio.com_2020}; thus, container resources are upper-bounded at 5\% of one CPU core, 20 MB of RAM, and 1 Mbps of bandwidth.

We measure accuracy and resource consumption at the level of individual peers to obtain results that do not depend on the number of co-deployed peers. Thus, each experiment involves one Leader and one Follower (of \adaptive or \static, respectively).
Cumulative resource consumption for multiple nodes can be derived by scaling the results proportionally. 
 
 
\subsection{RQ1 - Monitoring Accuracy}
We investigate the accuracy of the collected data for both \adaptive{} and \static{} considering synthetic indicators following different representative trends. 
More in detail, we created a probe reporting readings from such indicators. This allowed us to test the correctness of \adaptive's adaptive behavior and verify its effectiveness. 
We defined 5 scenarios (also referred to as time series in the following) mimicking different key cases for an indicator conventionally ranging between 0 and 1: 
\begin{enumerate}[leftmargin=*]
\item \emph{stable} is a time series representing a regular and stable, almost constant, trend. It is generated by alternating two close values (0.8 and 0.83), each of which remains stable for 14 seconds; 
\item \emph{unstable} is a time series that represents an irregular and erratic indicator with fluctuating values in the range $[0.5,0.85]$; a real-life trace on CPU utilization was used as a base.  
\item \emph{stable-unstable} is a time series that alternates phases of stability with phases of instability, with each phase lasting for about 150 seconds; 
\item \emph{random} is a time series with chaotic and totally unpredictable values; it is generated by a sequence of random values uniformly distributed in the range $[0,1]$; 
\item \emph{spiky} is a time series with mostly regular values interleaved with rare spikes; it is generated by alternating stable values for 28 seconds, unstable values for 12 seconds, and then a spike value for 4 seconds. 
\end{enumerate}
Every time series has a duration of 10 minutes, except for the stable-unstable time series that lasts 20 minutes since it is a combination of the stable and unstable time series.

\static and \adaptive's Follower peers are configured to forward the average of the last 20 measurements to their respective Leaders at each probing point. The \static Follower probes a new value from the monitored metric at a fixed interval: every 30 seconds. 
\adaptive exploits the \emph{\adaptRate} countermeasure to adjust the sampling rate to handle the variability of the monitored indicator. The hypothesis to test is that this can lead to improved monitoring accuracy because the Leader should have access to a higher number of samples when the monitored indicator is highly dynamic and fewer samples in the presence of more stable indicators.

We investigate the capability of the monitoring system to reconstruct the shape of the monitored indicators at the level of both the Follower, which directly samples the indicator, and the Leader, which collects a sequence of average values. 
We use the \emph{Root Mean Square Error} (RMSE), which measures the differences between the original and the reconstructed indicator, as the primary quality metric. A smarter sampling strategy should achieve a lesser error.
To appreciate the activity of the peers in relation to the monitored indicator, we also gauge the \emph{messages/second} ratio, that is, the ratio of the messages sent by the Follower to the Leader.  
Finally, for the spiky time series, we also computed the \emph{percentage of detected spikes}, which measures the capability of a monitoring technique to spot rare but significant events. 

\begin{table}[t]
\centering
\resizebox{\linewidth}{!}{%
\begin{tabular}{@{}clcccc@{}}
\toprule
\textbf{Scenario} &
  \multicolumn{1}{c}{\textbf{Quality Metric}} &
  \textbf{\adaptive{}} &
  \textbf{\static{}} &
  \textbf{Abs} &
  \textbf{Rel} \\ \midrule
                                       & RMSE (Follower)      & 0.019 & 0.020   & \cellcolor[HTML]{CBFFCA}- 0.181 & \cellcolor[HTML]{CBFFCA}- 5.0 \%     \\ \cmidrule(l){2-6} 
                                       & RMSE (Leader)        & 6.696 & 8.200    & \cellcolor[HTML]{CBFFCA}- 1.504 & \cellcolor[HTML]{CBFFCA}- 18.3 \% \\ \cmidrule(l){2-6} 
\multirow{-3}{*}{\textit{Stable}} &
  Messages/second &
  0.027 m/s &
  0.040 m/s &
  \cellcolor[HTML]{CBFFCA}- 0.013 &
  \cellcolor[HTML]{CBFFCA}- 32.5 \% \\ \midrule
                                       & RMSE (Follower)      & 0.087 & 0.131  & \cellcolor[HTML]{CBFFCA}- 0.044 & \cellcolor[HTML]{CBFFCA}- 33.6 \% \\ \cmidrule(l){2-6} 
                                       & RMSE (Leader)        & 2.428 & 5.033  & \cellcolor[HTML]{CBFFCA}- 2.605 & \cellcolor[HTML]{CBFFCA}- 51.7 \% \\ \cmidrule(l){2-6} 
\multirow{-3}{*}{\textit{Unstable}} &
  Messages/second &
  0.217 m/s &
  0.037 m/s &
  \cellcolor[HTML]{FFCCC9}+ 0.180 &
  \cellcolor[HTML]{FFCCC9}+ 486.5 \% \\ \midrule
                                       & RMSE (Follower)      & 0.108 & 0.122  & \cellcolor[HTML]{CBFFCA}- 0.014 & \cellcolor[HTML]{CBFFCA}- 11.5 \% \\ \cmidrule(l){2-6} 
                                       & RMSE (Leader)        & 5.269 & 6.546  & \cellcolor[HTML]{CBFFCA}- 1.277 & \cellcolor[HTML]{CBFFCA}- 19.5 \% \\ \cmidrule(l){2-6} 
\multirow{-3}{*}{\textit{Stable-unstable}} &
  Messages/second &
  0.103 m/s &
  0.035 m/s &
  \cellcolor[HTML]{FFCCC9}+ 0.068 &
  \cellcolor[HTML]{FFCCC9}+ 194.3 \% \\ \midrule
                                       & RMSE (Follower)      & 0.235 & 0.321  & \cellcolor[HTML]{CBFFCA}- 0.086 & \cellcolor[HTML]{CBFFCA}- 26.8 \% \\ \cmidrule(l){2-6} 
                                       & RMSE (Leader)        & 1.899 & 10.683 & \cellcolor[HTML]{CBFFCA}- 8.784 & \cellcolor[HTML]{CBFFCA}- 82.7 \% \\ \cmidrule(l){2-6} 
\multirow{-3}{*}{\textit{Random}} &
  Messages/second &
  0.217 m/s &
  0.037 m/s &
  \cellcolor[HTML]{FFCCC9}+ 0.180 &
  \cellcolor[HTML]{FFCCC9}+ 486.5 \% \\ \midrule
                                       & RMSE (Follower)      & 0.092 & 0.087  & \cellcolor[HTML]{FFCCC9}+ 0.005 & \cellcolor[HTML]{FFCCC9}+ 5.8 \%  \\ \cmidrule(l){2-6} 
                                       & RMSE (Leader)        & 5.713 & 6.251  & \cellcolor[HTML]{CBFFCA}- 0.538 & \cellcolor[HTML]{CBFFCA}- 8.6 \%  \\ \cmidrule(l){2-6} 
                                       & Messages/second      & 0.062 m/s & 0.037 m/s  & \cellcolor[HTML]{FFCCC9}+ 0.025 & \cellcolor[HTML]{FFCCC9}+ 67.6 \% \\ \cmidrule(l){2-6} 
\multirow{-4}{*}{\textit{Spiky}} & Detected spikes & 30 \%    & 0      & \cellcolor[HTML]{CBFFCA}+30 \%     & \cellcolor[HTML]{CBFFCA}{-}                                \\ \bottomrule
\end{tabular}%
}
\caption{Accuracy of \adaptive and \static for the 5 scenarios. Green (Red) cells indicate a better (worse) result obtained by \adaptive compared to the \static.}
\label{tab:rq1-results}
\end{table}

\paragraph{Results}
Table~\ref{tab:rq1-results} summarizes the results obtained by both \adaptive and \static for the considered 5 scenarios. 
The last two  columns show the absolute (Abs) and relative (Rel) deviations between the \adaptive and \static results for any of the presented quality indicators. Green (Red, respectively) cells indicate a better (worst) result obtained by \adaptive compared to the \static baseline.
We can observe that \adaptive estimates the observed indicator more accurately than \static at both levels of the Leader-Followers hierarchy in 4 out 5 scenarios (viz. \emph{stable}, \emph{unstable}, \emph{stable-unstable}, \emph{random}). The reduction in the RMSE reached $33.6\%$ at Follower level (\emph{unstable} scenario) and $82.7\%$ at Leader level (\emph{random} scenario).
A higher accuracy, however, comes at the cost of a higher number of messages exchanged in the 4 scenarios where the monitored indicator is more erratic (nearly 5 times more than \static in the worst case), and fewer messages produced when the indicator is stable (saving nearly one third of the messages). 
These results demonstrate the capability of \adaptive to modulate resource consumption as needed, exchanging more messages only if the monitored indicator requires finer sampling, saving bandwidth otherwise.



\begin{figure*}[!ht]
    \centering
    \includegraphics[width=\linewidth,height=4cm]{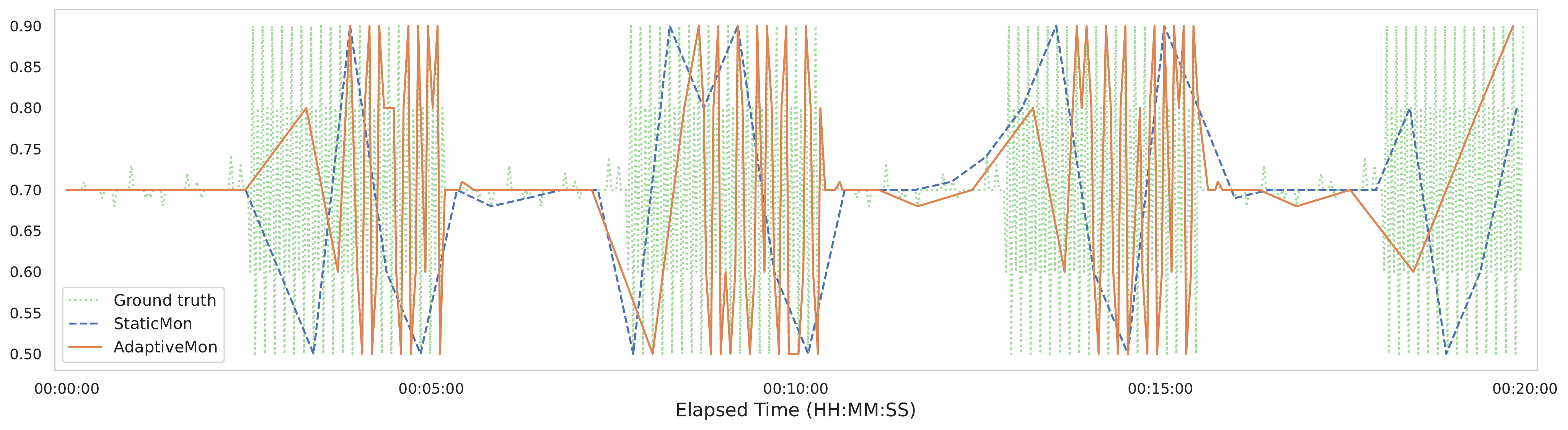}
    \caption{\adaptive and \static Follower time series estimations for the \emph{stable-unstable} scenario.}
    \label{fig:stable-unstable-follower}
\end{figure*}

\begin{figure*}[!ht]
    \centering
    \includegraphics[width=\linewidth,height=4cm]{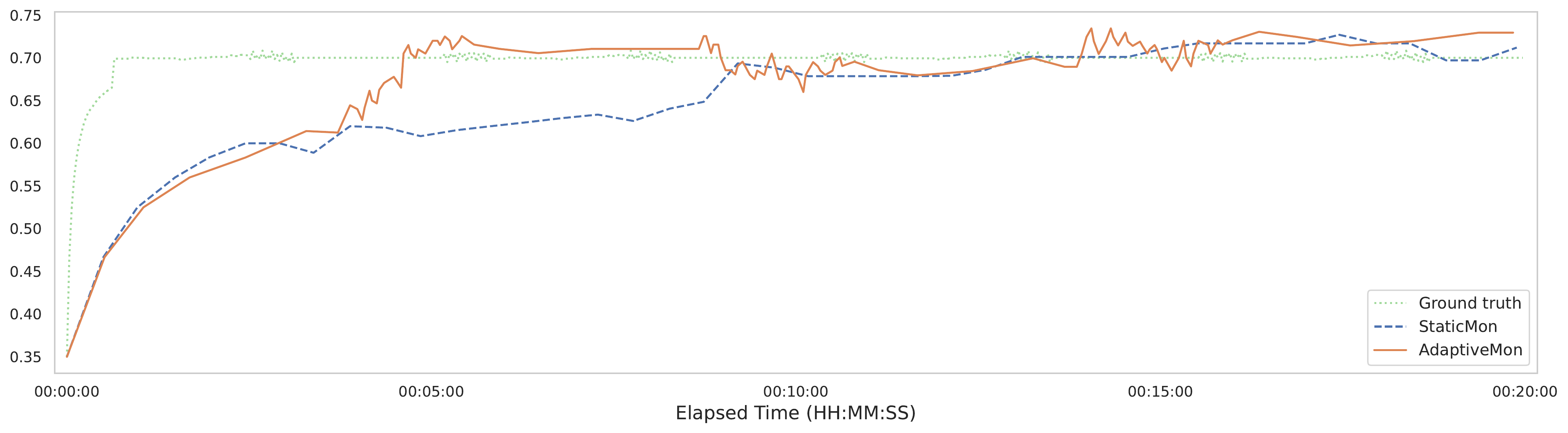}
    \caption{\adaptive and \static Leader time series estimations for the \emph{stable-unstable} scenario.}
    \label{fig:stable-unstable-leader}
\end{figure*}

Figures~\ref{fig:stable-unstable-follower} and~\ref{fig:stable-unstable-leader} exemplify the results of the comparison between \adaptive and \static for the \emph{stable-unstable} scenario. We can observe that the time series reconstructed by \adaptive (solid orange line) is closer to the reference indicator (dotted green line) than the \static baseline (dashed blue line). 
It is also interesting to notice how the rapid change in the observed trend is not immediately handled by \adaptive, which shows some delay in sensing the drift and adjusting the sampling rate. In contrast, \static always fails to follow the observed time series, confirming the importance of adaptivity in similar contexts.



\begin{figure*}[!ht]
    \centering
    \includegraphics[width=\linewidth,height=4cm]{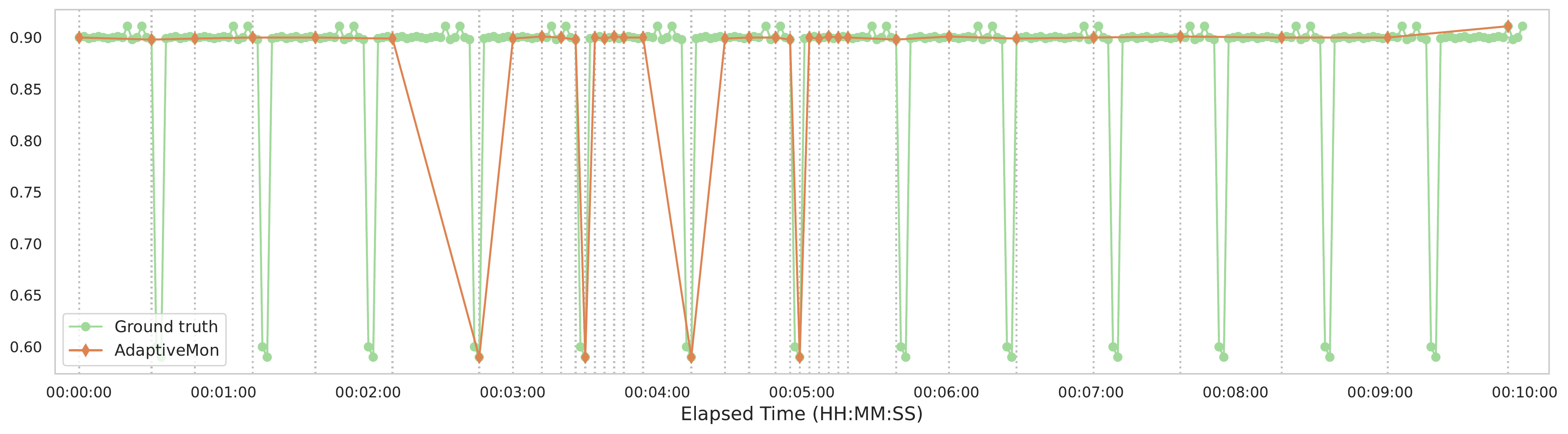}
    \caption{\adaptive Follower time series estimation for the \emph{spiky} scenario. The vertical dotted grey lines indicate the \emph{sampling rate}.}
    \label{fig:spikes-follower-sampling}
\end{figure*}

The \emph{spiky} scenario is the only one resulting in an increment of the RMSE metric for the adaptive Follower ($+5.8\%$). However, this increment is a consequence of the capability to (partially) follow the trend of the indicator. In fact, the \static configuration could detect spikes only incidentally, while \adaptive could change its sampling rate to increase the chance to capture them. Figure~\ref{fig:spikes-follower-sampling} illustrates a representative execution of  \static and \adaptive for the spiky scenario, with some spikes successfully detected by \adaptive only. Although successfully capturing some spikes, the reconstructed time series generates a higher error compared to the flat time series reconstructed by \static.
Indeed, this is a challenging scenario for both approaches (rare short events are hard to detect by monitoring techniques), and more work is required to design cost-effective monitoring techniques that can accurately address spikes. 



\subsection{RQ2 - Resource Consumption}
RQ2 investigates resource consumption considering the two countermeasures currently implemented in \adaptive, both in isolation and jointly. Again, \static{} is used as the baseline for the comparison. 
For the experiments reported in this section, we defined a probe (at the follower level only) that exploits the Docker Engine API\footnote{https://docs.docker.com/engine/reference/commandline/stats/} and PowerTOP\footnote{https://github.com/fenrus75/powertop} to collect the following quality metrics:

\begin{itemize}[leftmargin=*]
        \item CPU and memory consumption: the percentage of the host's CPU and memory used.
        \item NET I/O (MB): the cumulative amount of data 
        sent and received over its network interface from the beginning of the experiment.
        \item PIDs: the number of processes or threads spawned by the peer.
        \item PW (mW): the estimated instantaneous power consumption.
\end{itemize}


The three \adaptive configurations assessed in this RQ exploit the two strategies defined in Section~\ref{sec:approach}. \emph{\adaptRate} adjusts the sampling and forwarding rates of all the collected indicators from 30 to 60 seconds based on the monitored values.
\emph{\adaptMetrics} disables the collection of all indicators 
\changetxt{except of} power consumption if the battery level drops below a threshold.
\emph{\adaptCombined} uses both strategies. 
We study the impact of these configurations, along with the \static{} baseline, on resource consumption: a total of four possible configurations are therefore considered. In each experiment, the Follower peer is configured to collect the indicators from its node for 30 minutes and to apply the countermeasures at the beginning of the execution, in such a way the impact of the countermeasures can be accurately measured (in fact we collected more than 300 hundreds samples per metric).  

\paragraph{Results}
\begin{figure*}[!t]
    \centering
    \includegraphics[width=\linewidth, height=4.5cm]{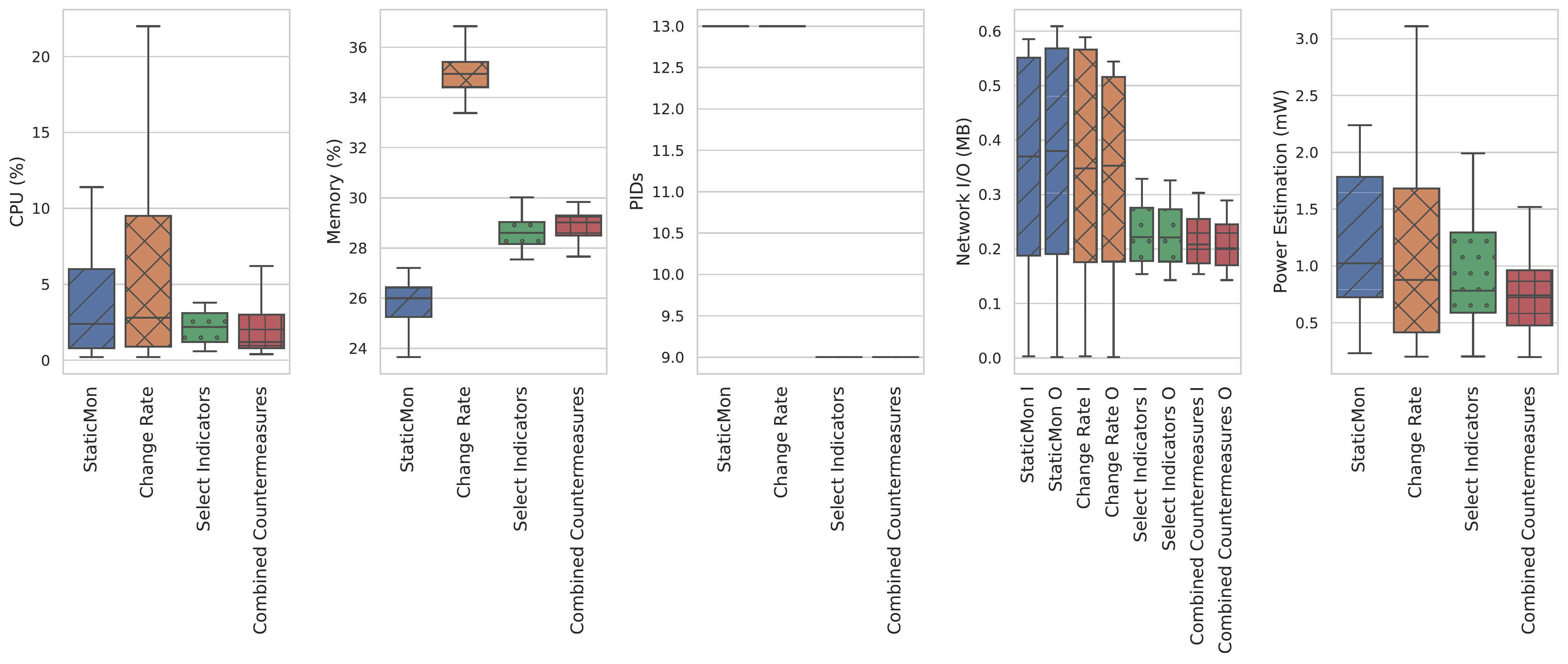}
    \caption{\static{} compared with \adaptive{} countermeasures for each of the collected quality metrics.}
    \label{fig:all-metrics-boxplots}
\end{figure*}

Figure~\ref{fig:all-metrics-boxplots} shows a series of five box plots (one for each quality metric) where each plot compares the four compared configurations visually.

We checked the significance of the differences between distributions with the non-parametric Mann–Whitney $U$ test~\cite{mann1947test}, as we observed (via the Shapiro-Wilk test~\cite{shapiro1965analysis} ) that such differences are not normally distributed. We specifically checked if the observed differences between the baseline and any other configuration are statistically significant and if \emph{\adaptCombined} is significantly better than the individual adaptations strategies (\emph{\adaptRate} and \emph{\adaptMetrics}). 
We considered a significance level $\alpha = 0.05$, and we also computed the effect size of the observed phenomenon using the Wendt's formula~\cite{wendt1972dealing}. Table~\ref{tab:statistical-tests} shows the significant cases only with their corresponding effect size using the conventional categories \textit{small} (less than 0.3), \textit{medium} (between 0.3 and 0.5), and \textit{large} (greater than 0.5).


\begin{table}[!t]
\centering
\resizebox{\linewidth}{!}{%
\begin{tabular}{@{}lll@{}}
\toprule
\textbf{Quality Metric}                      & \textbf{Comparison}       & \textbf{Effect Size} \\ \midrule
\multirow{2}{*}{\textit{CPU consumption}}          & \adaptRate vs \adaptCombined & small (0.197)        \\
 & \adaptMetrics vs \adaptCombined & small (0.260)  \\ \midrule
\multirow{5}{*}{\textit{Memory consumption}}       & \static vs \adaptMetrics        & large (0.858)        \\
 & \static vs \adaptRate            & large (0.993)   \\
 & \static vs \adaptCombined         & large (0.861)  \\
 & \adaptRate vs \adaptCombined   & large (0.990)  \\
 & \adaptMetrics vs \adaptCombined & small (0.279)  \\ \midrule
\multirow{4}{*}{\textit{\# spawned sub-processes}} & \static vs \adaptMetrics        & large (0.980)        \\
 & \static vs \adaptRate            & small (0.027)  \\
 & \static vs \adaptCombined         & large (0.963)  \\
 & \adaptRate vs \adaptCombined   & large (0.962)  \\ \midrule
\multirow{5}{*}{\textit{Network Input}}            & \static vs \adaptMetrics        & medium (0.449)       \\
 & \static vs \adaptRate            & small (0.101)  \\
 & \static vs \adaptCombined         & medium (0.483) \\
 & \adaptRate vs \adaptCombined   & medium (0.431) \\
 & \adaptMetrics vs \adaptCombined & small (0.145)  \\ \midrule
\multirow{5}{*}{\textit{Network Output}}                    & \static vs \adaptMetrics        & medium (0.473)       \\
 & \static vs \adaptRate            & small (0.245)  \\
 & \static vs \adaptCombined         & large (0.521)  \\
 & \adaptRate vs \adaptCombined   & medium (0.460) \\
 & \adaptMetrics vs \adaptCombined & small (0.203)  \\ \midrule
\textit{Battery power estimation}                  & \static vs \adaptCombined        & medium (0.376)       \\ \bottomrule
\end{tabular}%
}
\caption{Statistically valid comparisons for all the quality metrics with their associated effect size.}
\label{tab:statistical-tests}
\end{table}

We only observe marginal differences in CPU consumption. In particular, differences between \adaptive and \static are not significant, and \emph{\adaptCombined} introduces significant but small differences compared to employing the other two adaptive strategies individually.  

The memory consumption results show statistical significance for all cases with a large effect size for all comparisons, except for \emph{\adaptMetrics} compared to \emph{\adaptCombined} where the effect size is small.  
The impact of the adaptive strategies on the memory indicator is antithetic: while we notice an increase in memory consumption of about 10\% for \emph{\adaptRate} (compared to \static), memory overhead decreases to about 3\% when \emph{\adaptMetrics} or \emph{\adaptCombined} are used. 
These results can be easily explained considering that although the MAPE-K control loop increases the amount of memory used by \adaptive when all probes are active, while when these are disabled (freeing the associate resources) the overall average memory usage decreases.   


Since limiting the number of processes can be particularly important when the underlying device platform is resource constrained we measured the number of sub-processed spawned by all compared configuration. In this regard, the number of spawned sub-processes show significant reduction with large effect size when \emph{\adaptMetrics} and \emph{\adaptCombined} are used.

\begin{figure}[!t]
    \centering
    \includegraphics[width=\linewidth]{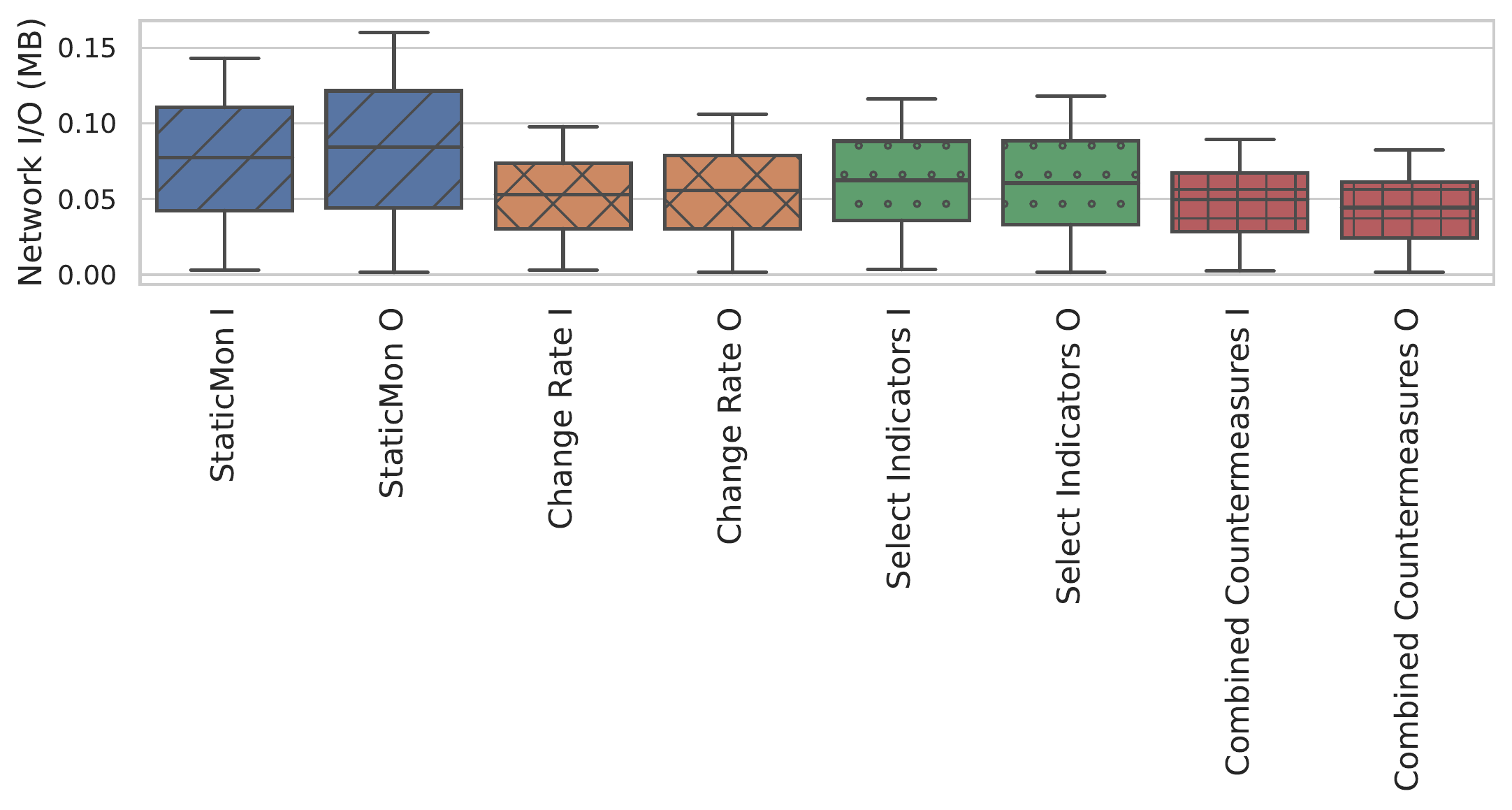}
    \caption{\static{} compared with \adaptive{} countermeasures for the network I/O metrics when the bandwidth is not measured by the Follower.}
    \label{fig:all-metrics-boxplots-wb}
\end{figure}

Limiting bandwidth consumption is also extremely important in fog environments, especially in the edge. As a matter of fact, limiting I/O operations is crucial when the network bandwidth is limited and shared by multiple devices and thus can be quickly saturated. Moreover, intensive communication implies high power consumption, a threat to energy efficiency and batteries lifespan in portable devices.
The results for network I/O show statistically significant reduction for all adaptive configurations compared to \static, with an effect size ranging from small to medium. Note that results for Input (I) and Output (O) present a similar behavior for the same configuration.   
More in detail, results show a small effect size for \emph{\adaptRate} versus \static comparison. Since we expected a stronger impact of rate adaptation in this context,  we analyzed the behavior of the probes, and discovered that the bandwidth gauge exploits iPerf\footnote{https://iperf.fr/}, which measures the bandwidth by saturating it with packets. Such an invasive behavior nullifies the potential benefits of a dynamic sampling rate. 
Therefore, to further investigate this dimension, we repeated the experiments by disabling the bandwidth monitoring probe for both \static{} and \adaptive. Results are presented in Figure~\ref{fig:all-metrics-boxplots-wb}. The impact of \emph{\adaptRate} is now remarkable, with a reduction on transmitted data ranging between 31\% and 34\%, with an even higher reduction (between the 37\% and the 49\%) when both countermeasures are simultaneously active.

Finally, the results on power consumption show meaningful differences only for \static{} versus \emph{\adaptCombined}, suggesting that the individual countermeasures may introduce limited benefits. Still, their combination can significantly improve battery lifetime (with an estimated reduction in power usage of about 36\%).

In summary, \adaptive{} can help save network I/O and devices battery without impacting CPU utilization by paying for an extra memory consumption to store the data necessary to run the adaptive mechanisms. This result can be beneficial in the Fog since monitoring systems are called to reduce network overhead without impacting on power consumption~\cite{abderrahim2017holistic}, especially when devices are battery-powered. At the same time, even resource constrained devices at the edge of the network are usually well equipped in terms of memory, mitigating the impact of a limited extra memory consumed by the monitoring system.

\subsection{Threats to Validity}
Our study is affected by both internal and external threats to validity. The main internal threats to validity concern with the design of the scenarios used to study RQ1. We proposed five scenarios to mimic different trends. Although indicators collected in real scenarios may behave differently than the ones we investigated, the results obtained with our stereotyped trends are still informative, at least locally (e.g., it is possible to refer to the results reported in the paper for an indicator that becomes unstable or too high). 

The definition of the logical states for an indicator depends on several parameters, which are application-dependent. In this paper, we studied how \adaptive can be used to obtain self-adaptive capabilities relevant to monitoring, focusing on the assessment of simple countermeasures that do not strongly depend on the domain. Assessing \adaptive in the context of dedicated application scenarios is part of our future work and is out of the scope of this paper.

The generality of the results obtained about efficiency (RQ2) might depend on the specific implementation used and the size of the experiment. We used an independent implementation for \static (i.e., FogMon) to minimize any implementation bias, and we added the adaptive behavior to this implementation. To further reduce any implementation threat, we released our solution publicly. In principle, additional experiments may lead to different results. However, we obtained quite clear evidence and we checked the statistical significance of the results to mitigate the risk of overgeneralization. 


\section{Related Work}\label{sec:related-work}

In the last decade, a large number of monitoring solutions, both commercial and academic, have been proposed for the Cloud~\cite{calero2014monpaas, tundo2019varys, trihinas2014jcatascopia, povedano2013dargos, olups2016zabbix, barth2008nagios, turnbull2018monitoring, shatnawi2018cloudhealth, elasticstack_2022}. However, they are seriously challenged by several characteristics of the Fog, such as its massively distributed infrastructure characterized by frequent changes to the topology and the presence of heterogeneous and resource constrained devices~\cite{taherizadeh2018monitoring, yousefpour2019all}.

A recent study by Taherizadeh et al.~\cite{taherizadeh2018monitoring} investigated the requirements that must be satisfied by monitoring platforms specialized for adaptive applications orchestrated upon the Cloud-to-Thing continuum. It is apparent from this survey that none of the solutions currently available for the Cloud can satisfy all the requirements, identifying decentralization and resource optimization via self-adaptation as two of the main open challenges.
Similar conclusions have been reported by Abderrahim et al.~\cite{abderrahim2017holistic} who explicitly identify the adaptability of the granularity of the reported measurements as one of the key properties for a fog monitoring system.



Monitoring approaches specifically designed for the fog environment have been recently investigated~\cite{brandon2018fmone, grossmann2017monitoring, souza2018osmotic, forti2021lightweight}. 
In particular, FMonE~\cite{brandon2018fmone} is a monitoring tool that relies on a container orchestration system to build monitoring pipelines, addressing the distinctive features of a fog infrastructure. It provides users with the flexibility to define their monitoring pipelines and operate them across the active regions. The container orchestration system is the (centralized) module of the FMonE architecture responsible for launching the monitoring agents and the data backends in each region.

PyMon~\cite{grossmann2017monitoring} is a lightweight prototypical monitoring framework available for relevant Docker-enabled architectures such as ARM, AARCH64 and x86\_64, and particularly suitable for single board computers (SBC) at the edge of the network. It extends the host-based monitoring tool Monit with capabilities to inspect running Docker containers.

Souza et al.~\cite{souza2018osmotic} proposed a monitoring tool that extends the CLABS model~\cite{alhamazani2015cross} and is capable of monitoring resources by deploying services between the Edge and the Cloud.

Unlike \adaptive{}, none of these solutions implement adaptive policies to adapt the behavior of the monitored system to the collected  data. Furthermore, they are not based on a P2P architecture, so that, they struggle to cope with some of the fog distinctive traits such as heavily distributed infrastructures, rapid changes in the topology, and communication links failures.

FogMon~\cite{forti2021lightweight} is the fog-oriented monitoring framework that we extended to implement \adaptive{}. It collects and aggregates data about resource consumption, network conditions, and IoT devices connected to fog nodes. It exploits a two-tier (Leader-Follower) P2P architecture and gossip protocols to reduce the network overhead. Also, it adapts the number of Leader nodes in the P2P overlay and the underlying Followers topology based on current network conditions. However, it does not provide  self-adaptive behaviors to govern the internal functioning of the monitoring system as designed in \adaptive. For instance, FogMon cannot be used to dynamically change the set of the collected indicators or the sampling rate.

Among the works that are not specifically designed for the Fog, it is worth mentioning some that still relate to \adaptive. In particular, ADMin~\cite{trihinas2017admin} is an IoT-specific monitoring framework designed to reduce devices' energy consumption and the volume of data sent over the network. This is achieved essentially by adapting the rate at which devices disseminate monitoring streams based on run-time knowledge (e.g., stream evolution, variability, seasonality).

Also, Tangari et al.~\cite{tangari2018self} propose a self-adaptive and decentralized framework for resource monitoring in the scope of Software Defined Networks (SDN). It enables metrics collection through a self-tuning and adaptive monitoring technique that adjusts its settings based on traffic dynamics to balance operation costs with monitoring accuracy while reducing network overhead.
Compared to \adaptive, the proposed framework lacks generality since the adaptation capabilities are limited to some predetermined aspects and is not designed to support the capability to run multiple and diverse adaptation rules.

Finally, SkyEye~\cite{graffi2017skyeye} is a tree-based monitoring solution operating on structured P2P overlay networks. It provides continuous monitoring for a wide range of metrics for all peers in the network. It is characterized by a tree structure, which enables peer partitions in a hierarchical fashion. The aggregated statistics received by the upper layers of the tree describe the statistics of the peers in the corresponding sub-trees. Messages are used to disseminate the global statistics retrieved from the top levels and maintain the tree topology. However, unlike \adaptive it is not explicitly designed for the Fog, and it completely lacks adaptivity.

 \vfill
\section{Conclusions}\label{sec:conclusions}
Fog monitoring calls for approaches that can deal with the dynamism of the environment while taking the available resources into consideration. P2P monitoring approaches cope with some of the architectural-level traits of the Fog, such as nodes that can dynamically join and leave the network or malfunctioning in their communication links, but do not provide self-adaptive capabilities that can be used to dynamically adapt the behavior of the monitoring system to changing conditions.

In this paper, we presented \adaptive, a self-adaptive P2P monitoring solution that exploits a knowledge base continuously fed with monitoring data to dynamically change the system behavior by activating countermeasures. Experimental results show that adaptive behaviors can improve monitoring accuracy while optimizing the usage of the available resources, in comparison to a non-adaptive solution.

Future work will focus on extending the self-adaptive capabilities of \adaptive by considering several application scenarios, including anomaly detection and self-healing, and multiple application domains such as Service Level Agreements (SLAs) and Function-as-a-Service (FaaS) monitoring in the fog environment. We also plan to study how to exploit the MAPE-K paradigm at the level of the whole P2P monitoring system, to deliver adaptive capabilities at the system level (e.g., to run adaptation procedures that require the synchronized intervention of multiple peers), instead of the individual peers only.


\bibliographystyle{ACM-Reference-Format}
\bibliography{bibliography}

\end{document}